\documentclass{article}

\usepackage{arxiv}

\usepackage[utf8]{inputenc} 
\usepackage[T1]{fontenc}    
\usepackage{hyperref}       
\usepackage{url}            
\usepackage{booktabs}       
\usepackage{amsfonts}       
\usepackage{nicefrac}       
\usepackage{microtype}      
\usepackage{lipsum}		
\usepackage{graphicx}
\usepackage{titlesec}

\renewcommand{\headeright}{}
\renewcommand{\undertitle}{}

\title{Lyra - containerized microservices for browsing shared biomedical data}

\author{ Michael Huttner \\
	Institute of functional genomics \\
	University of Regensburg \\
	Regensburg, Germany\\
	\texttt{michael.huttner@ukr.de} \\
	\And
    Claudio Lottaz \\
	Institute of functional genomics \\
	University of Regensburg \\
	Regensburg, Germany\\
	\texttt{claudio.lottaz@ukr.de} \\
	\And
    Christian Kohler\\
	Institute of functional genomics \\
	University of Regensburg \\
	Regensburg, Germany\\
	\texttt{christian.kohler@ukr.de} \\
	\And
    Rainer Spang \\
	Institute of functional genomics \\
	University of Regensburg \\
	Regensburg, Germany\\
	\texttt{rainer.spang@ur.de} \\
}

\titlespacing*{\section}{0.2cm}{0.0cm}{0.0cm}
\begin{document}

\maketitle


\

\begin{abstract}

Research papers in the biomedical field come with large and complex data sets that
are shared with the scientific community as unstructured data files via public data 
repositories. Examples are sequencing, microarray, and mass spectroscopy data.
The papers discuss and visualize only a small part of the data, the part that is in
its research focus. For labs with similar but not identical research interests different parts of
the data might be important. They can thus download the full data, preprocess it, integrate it
with data from other publications and browse those parts that they are most interested in.
This requires substantial work as well as programming and analysis expertise that only 
few biological labs have on board.
In contrast, providing access to published data over web browsers makes all data
visible, allows for easy interaction with it, and lowers the barrier to working with 
data from others.

We have developed \textit{Lyra}, a collection of microservices that allows labs to 
make their data easily browsable over the web. Currently we provide tools for (a)
 insertion of genomic, proteomic, transcriptomic and metabolomic data, (b) 
 cross linking data from different  publications via automatic conversion of 
 over 200 molecular identifier types, (c) fast data access and search over a JSON API, 
and (d)  dynamic and interactive visualization in the users web browser.

\end{abstract}

\section{Microservices in containers}
\begin{figure*}[b]
\centering
\includegraphics[width=0.8\textwidth, trim={6cm 6cm 6cm 4cm}]{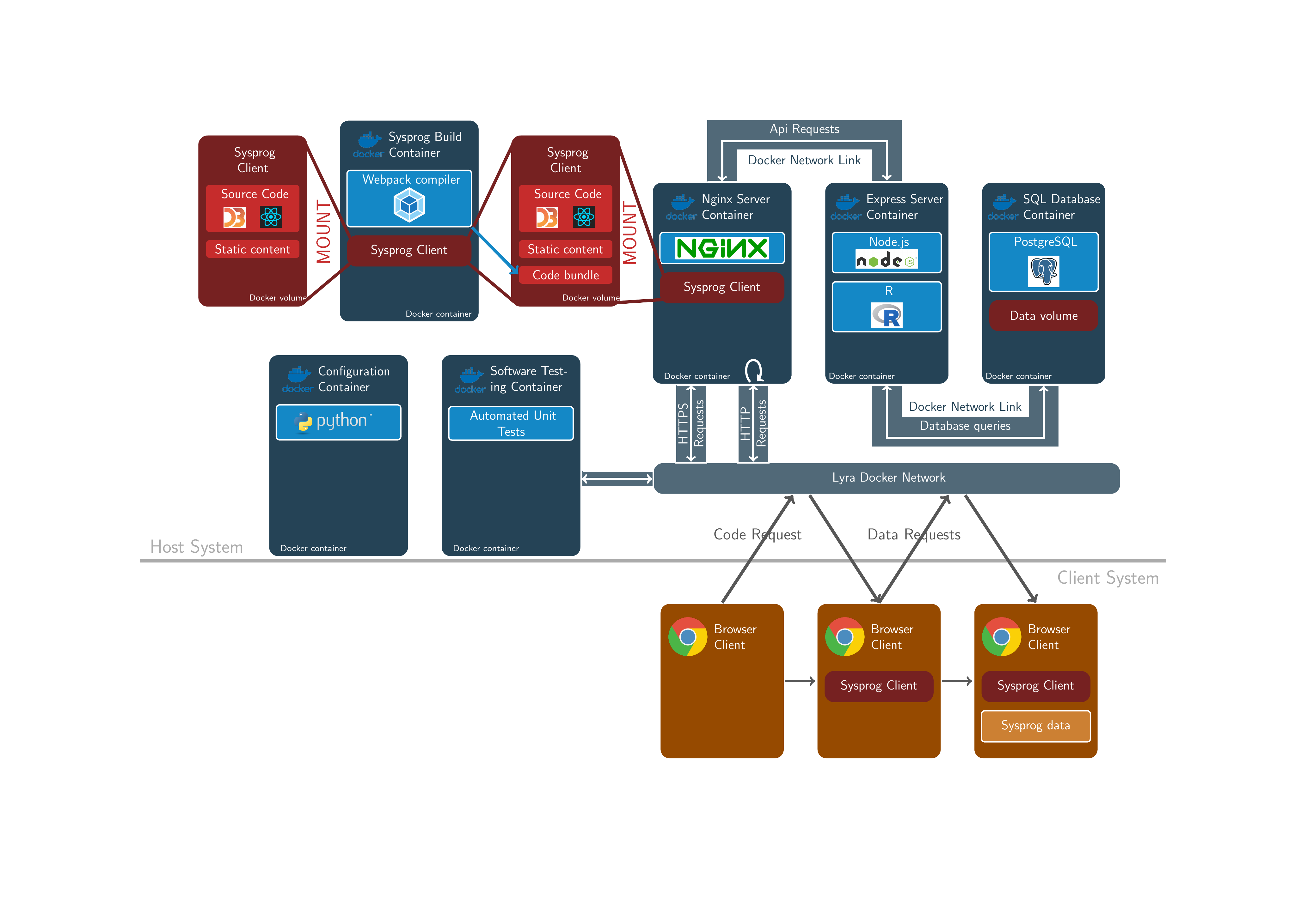}
\caption{
    Complete structure of the Lyra application, with one of the currently implemented Clients \textit{Sysprog}.
    We use three main containers, the Nginx static server, the NodeJS Express API server, and a PostgreSQL Database.
}
\label{fig:dockerstructure}
\end{figure*}

Key to our application design is a strict separation of software components through isolating them 
into distinct microservices running in containers (Figure \ref{fig:dockerstructure}).
Lyra is thus independent of the underlying hardware, leaving only one dependency: that on
 \textit{Docker}. In fact, Lyra can be deployed with minimal effort, or even
deployed automatically  to modern compute clusters, such as kubernetes.
Moreover, Lyra is built to scale.  The API server process is stateless and can be scaled by replication, 
given proper load balancing. 
Lyra uses a \textit{PostgreSQL} database with proven reliabilty and performance in enterprise
clustering solutions. Since our use case is heavily biased towards database reads over writes, creating read-only 
database replicas is simple and scales well. This allows Lyra to provide direct, fast and properly indexed 
access to its data.

\section{Knowledge graph}
\begin{figure*}[b]
\centering
\includegraphics[width=0.7\textwidth, trim={6cm 10cm 6cm 3cm}]{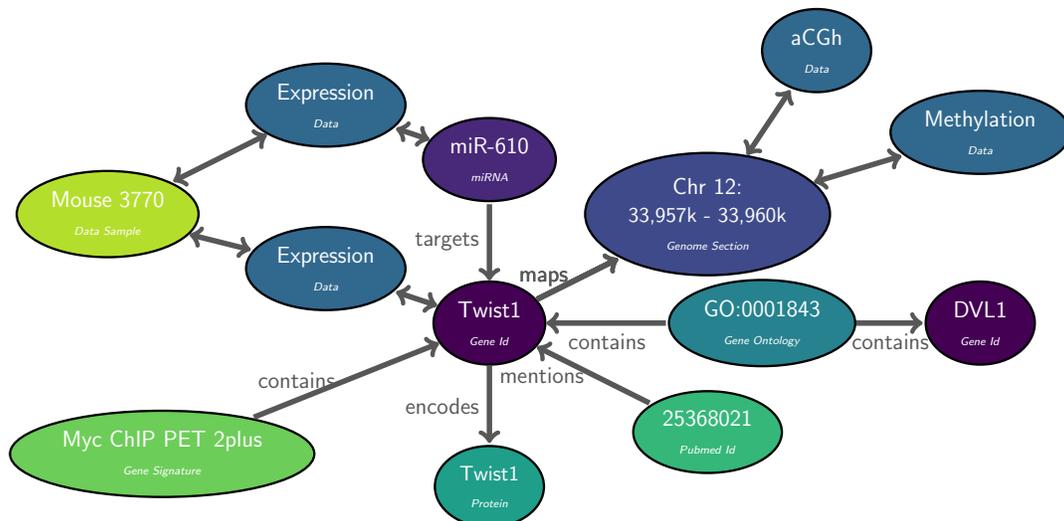}
\caption{
    Representation of the \textit{Lyra} knowledge graph, we represent any identifier as a Node in the graph and
    encode any relationships as edge. Relationships can be derived from existing annotation, 
    or based on custom data e.g correlation. Links to available data are also encoded as edges in the graph.
}
\label{fig:knowledgegraph}
\end{figure*}

Our typical user wants to see the data from one publication in the context of data from other publications.
Thus this data has to be cross linked and Lyra does a lot of this work fully automatic.
Central to that is the Lyra \textit{knowledge graph} (Figure \ref{fig:knowledgegraph}).  
We define the set of nodes of this graph as any possible identifier, e.g a gene name
a sample id, an ensembl id, etc. The set of edges can be any relationship between
nodes. For example:  The human gene name \texttt{TWIST1} corresponds to the id \texttt{ENSG00000122691}
in the ensemble database, for this ID Lyra has expression data from 3 publications, its mouse homolog is
\textit{Twist1} for which Lyra has expression data from one more publication. 
Moreover, \texttt{TWIST1} is located on chromosome 12 at position $33,957k$ - $33,960k$, a region for which Lyra
 has both aCGH and DNA-Methylation data, and so on.
 Once a new data set is loaded to Lyra we check automatically for known identifiers and add corresponding edges
 to the knowledge graph. The front-end can then automatically pull related information across data sets and
 make them jointly visible.

 The knowledge graph is also the core of the search functionality, similar to the \textit{google knowledge graph}. We build an index 
of all nodes in the graph and make it searchable this way. A data user can do the following
\begin{enumerate}
    \setlength\itemsep{0.1em}
    \item Start with a search query: Let's say a gene name
    \item We present matching identifiers from the knowledge graph 
    \item The user selects a match, which is a node somewhere in the graph
    \item We find relevant data by finding the \textit{shortest path} in the graph to a data node.
    \item We can also suggest other related identifiers just by finding neighboring nodes in the graph.
\end{enumerate}
The only required input is a single keyword, the rest happens in the background.

A data provider has two options: (a) he can set up a local Lyra server, which generates a browsable
webpage for only the data in his publication, (b) he can link his data to existing Lyra serves. By doing so,
the knowledge graph enables the data providers to make their data 
discoverable for researches that browse a different but related data set. This is done by linking the data 
into the graph and to existing identifiers, either automatically  or via custom links that can be uploaded 
as  edges in the knowledge graph.

\section{Visualization}
Based on this fast data access we created a client library for the javascript framework
\textit{React} and currently maintain two web clients based on this library, 
\textit{Lymmml} focusing on human lymphoma, and \textit{Sysprog} for mouse breast
cancer models. 

\begin{figure}[t]
\centering
\includegraphics[width=0.4\textwidth]{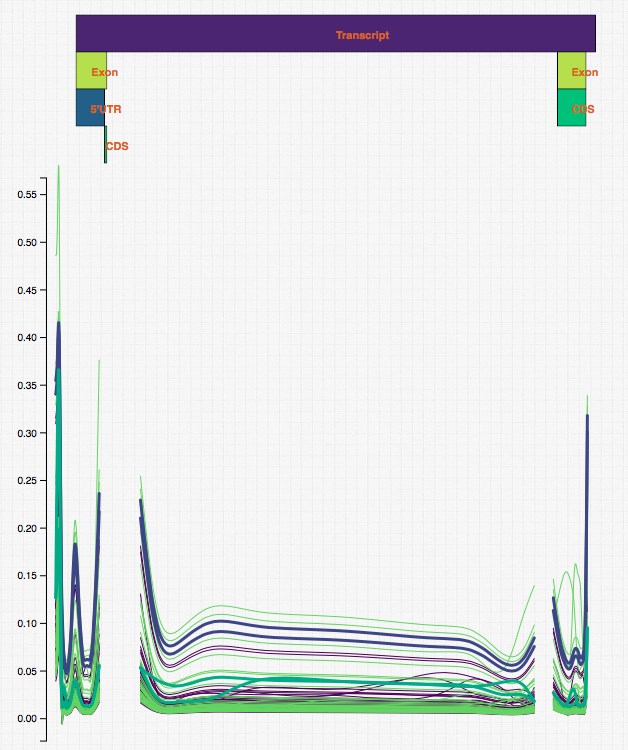}
\caption{
    Example visualization of DNA methylation data, the plot transforms using a \textit{fisheye}
    projection, magnifying the part the user selects.
}
\label{fig:methylation}
\end{figure}

Each of the clients contains custom svg data visualizations, based on \textit{d3.js} and
we implemented a flexible visualization pipeline for the client library to allow developers to 
quickly create custom visualization that fit the data and the problem domain.
 An Example can be seen in Figure \ref{fig:methylation}.

\section{Conclusion}

We provide software for data sharing that is  highly flexible and robust, and ready for cloud computing.
 The split into many loosely coupled microservices
allows others to take and adapt just the parts they need, and integrate them with already existing solutions.
In the future we will provide full integration for \textit{kubernetes} the leading container orchestration
platform, making it even easier to deploy \textit{Lyra} on bare-metal compute clusters or in commercial
cloud solutions.

\section{Availability}
\textit{Lymmml} is available at \url{lymmml.spang-lab.de} with many public datasets. \\[0.3cm]
\textit{Sysprog} is available at \url{sysprog.spang-lab.de}, you may request access by sending an email to \texttt{sekretariat.genomik@ukr.de}
\subsection{Source code}
The Lyra source code is available on github, in the following github repositories:
\begin{itemize}
    \item[--] \url{https://github.com/mhuttner/lyra-api-server} for the lyra API server. 
    \item[--] \url{https://github.com/mhuttner/lyra-lymmml-client} for the lymmml react client. 
    \item[--] \url{https://github.com/mhuttner/lyra-client-library} a javascript library for lyra clients. 
\end{itemize}

\end{document}